
\input phyzzx
\hfill
{KHTP-93-07 \par}
\vskip-15pt
\hfill
{SNUCTP-93-37 \par}
\title{\seventeenbf  Solutions of  Conformal Turbulence on a Half Plane}
\author{B.K. Chung, Soonkeon Nam,
\foot{E-mail address: nam@nms.kyunghee.ac.kr}
Q-Han Park,
\foot{E-mail address: qpark@nms.kyunghee.ac.kr}
and H.J. Shin
\foot{E-mail address: shin@SHIN.kyunghee.ac.kr} }
\address{\it  Department of Physics\break
	and\break
       Research Institute for Basic Sciences\break
	 Kyung Hee University\break
	  Seoul, 130-701, Korea}
\abstract{
Exact solutions of conformal turbulence restricted on a upper half plane are
obtained. We show that the inertial range of  homogeneous and   isotropic
turbulence with constant enstrophy flux develops in a distant region
from the boundary. Thus in the presence of  an anisotropic boundary,
these exact solutions of turbulence  generalize Kolmogorov's solution
consistently and differ from the Polyakov's bulk case which requires a
fine tunning of coefficients. The simplest solution in our case is given
by the minimal model of $p=2, q=33$ and moreover we find a fixed point
of solutions when $p,q$ become large.
}
\endpage
\def\zb{{\overline z}}


\REF\book{A.S. Monin and A.M. Yaglom, Statistical Fluid
Mechanics, The MIT Press, Cambridge (1975).}
\REF\PolyI{A.M. Polyakov, Princeton Univ. preprint PUPT-1341,
hep-th/9209046.}
\REF\PolyII{A.M. Polyakov, Nucl. Phys. {\bf B396} (1993) 367.
}
\REF\turbexp{B. Legras, P. Santangelo, and R. Benzi, Europhys. Lett. {\bf 5}
(1988)
37. }
\REF\Kol{A.N. Kolmogorov, J. Fluid Mech. {\bf 13} (1962) 82.}
\REF\Ferretti{G. Ferretti and Z. Yang, Rochester preprint UR-1296,
hep-th/9212021.}
\REF\Low{D.A. Lowe, Mod. Phys. Lett. {\bf A8} (1993) 923.}
\REF\Falkovich{G. Falkovich and A. Hanany, WIS-92/88/Nov-PH, hep-th/9212015. }
\REF\Matsuo{Y. Matsuo, Mod. Phys. Lett. {\bf A8} (1993) 619.}
\REF\CardyI{J.L. Cardy, Nucl. Phys. {\bf B240} [FS12] (1984) 514.}
\REF\KyungheeI{B.K. Chung, S. Nam, Q-H. Park, and H.J. Shin,
Phys. Lett. {\bf 309B} (1993) 58.}
\REF\KyungheeII{B.K. Chung, S. Nam, Q-H. Park, and H.J. Shin, Kyung Hee
preprint
KHTP-93-08.}
\REF\Kra{R. Kraichnan, Phys. Fluid, {\bf 10} (1967) 1417.}
In the theory of turbulence, Kolmogorov's theory of the inertial range
provides a  simple but powerful tool  in obtaining the scaling laws of
locally isotropic turbulence.\refmark\book\  Nevertheless, the microscopic,
field
theoretic foundation of the scaling law   so far has been  lacking.
Recently, Polyakov has proposed a completely new approach to this problem
based on the assumption of conformal invariance of velocity correlators
in the case of two dimensional fully developed turbulence.\refmark{\PolyI ,
\PolyII}
He  took  the
non-unitary minimal conformal field theory as a field theoretic framework
for turbulence and obtained exact solutions of turbulence with specific
scaling behaviors, different from the Kolmogorov's one, which seem to fit
better with experimental results.\refmark\turbexp\ However, due to the nature
of
non-unitarity,  one point functions of operators in the theory  do not
necessarily vanish unlike the unitary case. The arbitrariness of one point
 functions in the Polyakov's approach  makes the theory largely
undetermined and remains as an open problem.

In this letter,  we show that one point functions are determined  by
boundary conditions, as suggested by Polyakov, in particular  when they are
confined on a upper half plane. This allows us to  give a field theoretic
account for the Kolmogorov's hypothesis of inertial range.\refmark\Kol\
 We show that
the inertial range of  homogeneous and   isotropic  turbulence with constant
enstrophy flux develops in a distant region  from the boundary. Moreover we
argue that  exact solutions obtained by  Polyakov and others\refmark{\Ferretti
-
\Matsuo}
 for the  bulk
case  require an unnatural fine tunning of coefficients. Without  such a
fine tunning, we obtained series of solutions of conformal turbulence which
are different from the bulk case. The simplest solution in our case is
given by the minimal model of $p=2, q=33$. More interestingly, we find
a fixed point of conformal solutions when  $p,q$ become very large whose
energy spectrum exponent is $ -67/15\sim - 4.4667$.

The basic assumption of Polyakov's conformal  turbulence is to identify the
stream function $\psi $, arising from the velocity vector $v_{i}$ of
incompressible fluid such that $v_{i} = \epsilon_{ij}\partial_{j}\psi $, with a
certain primary operator of a non-unitary minimal model which, in the
inviscid and static case, satisfies the Navier-Stokes equation as well as
 the constant enstrophy condition.
The use of two dimensional conformal field theory(2-d CFT) allows us to
compute  exactly the velocity correlation functions, thereby the kinetic
energy of  fluid motion. In particular, the velocity two point function
 can be obtained directly from the stream two point function by
 differentiation,
 for example, $\langle v_{x}(z_{1})v_{x}(z_{2})\rangle =
 \partial_{y_{1}}\partial_{y_{2}}\langle \psi (z_{1})\psi (z_{2})\rangle
; z = x + iy $.

In principle, the two point function of stream function can be obtained
exactly.
However, in practice, for the brevity of our discussion and
 also in order to avoid the difficulty in solving for exact solutions, we
consider only
the short distance behavior  of  the velocity two point function which can be
obtained easily from the operator product expansion (OPE) of two
$\psi$'s;  $$\langle \psi (z_{1})\psi (z_{2}) \rangle
= \sum_{i} C_{\psi \psi }^{\phi_{i}} { \langle \phi_{i} \rangle \over
|z_{1}-z_{2}|^{4\Delta_{\psi }-2\Delta_{\phi_{i }}}} +{\rm descendents}
,\eqn\opeII$$
where $C_{\psi \psi }^{\phi_{i}} $ are structure constants and $\phi_{i}$
are primary operators of conformal dimension $\Delta_{\phi_{i}}$. The one
point functions $\langle \phi_{i} \rangle $ except for that of
the identity operator vanish identically when the vacuum is invariant under
$SL(2,C)$.  However, in the presence of large scale structures of scale $L$,
so that the vacuum is no longer invariant under $SL(2,C)$, the one point
function behaves as
$$ \langle \phi_{i} \rangle \sim C_{i}L^{-2\Delta_{\phi _{i}}}  \ \ . \eqn
\lone$$
For the unitary case, $\Delta_{\phi _{i}} \geq 0 $
and $\langle \phi_{i} \rangle $ approaches zero when $L$ becomes large.
However this  is no longer true for the non-unitary model where
$\Delta_{\phi _{i}}$ could be negative. Moreover  non zero values of one
point functions modify significantly other correlation functions so that
they become  essential ingredients in defining a non-unitary field theory.

In order to understand the role of the one point function and its
physical implications more clearly,
we first note that  they can be determined exactly by a specific boundary
condition.\refmark\KyungheeI\  Consider, for example, the two dimensional
turbulence restricted
on a upper half plane; Im $z = y >0$. A typical boundary condition is such
that the normal component of the velocity vector vanishes at the boundary.
In order to maintain the spirit of 2-d CFT approach to turbulence on a upper
half plane, we assume that the conformal symmetry of correlation functions
persists in the presence of boundary, but now in a reduced form which
preserves the boundary condition.  This is tantamount to reducing $SL(2,C)$
to $SL(2,R)$ in the case of the global conformal transformation  so that only
real translations and dilatations are involved. The domain under
consideration, the upper half plane, is invariant under global real
conformal transformations of $SL(2,R)/Z_{2}$ with an added point at
infinity. This boundary condition also requires correlation functions to
be invariant under $SL(2,R)$.  In the case of two-point function,  one could
obtain this by forming an $SL(2,R)$ invariant
$u$ $$u = { (z_{1} - \bar{z}_{1}) (z_{2} - \bar{z}_{2})
\over  (z_{1} - \bar{z}_{2}) (z_{2} - \bar{z}_{1})}, \eqn \slin$$ where two
points $z_{1}$ and $z_{2}$ are given in the upper half plane and
$\bar{z}_{1}$ and $\bar{z}_{2}$ are their images in the lower half plane.
In general, it is possible to show  that for the degenerate conformal field
theories, the $n$-point function
$\langle \phi (z_{1},\zb_{1})\cdots\phi (z_{n},\zb_{n})  \rangle_{b}$ can be
obtained systematically from the bulk $2n$-point function,
where $\langle\cdots\rangle_{b}$ denotes correlation functions in the upper
half plane.\refmark\CardyI\ This is so because the $n$-point function
$\langle \phi (z_{1},\zb_{1}) \cdots \phi (z_{n},\zb_{n})\rangle_{b}$
satisfies the same differential equation as does the bulk $2n$ point
function consisting of charges in the upper half plane as well as their
images in the lower half plane. In particular, one point functions on a
upper half plane can be obtained exactly from the known bulk two point
functions  such that
$$ \langle \phi_{i}(z) \rangle_{b} = d_{\phi_{i}}y^{-2\Delta_{\phi _{i}}}
; \ \  y ={\rm Im} z, \eqn
\onept$$
where $  d_{\phi_{i}} $ are arbitrary constants. These  constants
$  d_{\phi_{i}} $, as well
as the negative energy states $( \Delta_{\phi_{i}} < 0)$ in non-unitary
minimal models, seem to be unphysical at first sight. However,
they acquire physical meaning when the non-unitary model is  applied for an
effective description of an open physical system. In our case,
$  d_{\phi_{i}} $  obtain physical meaning through the velocity
correlators and become physical parameters of turbulence.
{}From Eq.\onept  , for example, we have
an average velocity profile; $$\langle v_{y} \rangle_{b}  = - \partial_{x}
\langle \psi \rangle_{b} = 0; \
\langle v_{x} \rangle_{b}  = \partial_{y}
\langle \psi \rangle_{b} = - 2 d_{\psi } \Delta_{\psi }y^{-2
\Delta_{\psi }-1}, \eqn \prof $$
where $d_{\psi }$ is a parameter which controls
the magnitude of average horizontal velocity.
Another example where $  d_{\phi_{i}} $  manifest physical meaning is
the kinetic energy of turbulent fluid. It is commonly expressed in terms of
the energy density at  point $(x,y)$  given by
$${1\over 2} \langle v_{\alpha}^{2}(x,y)\rangle_{b} = \int dk
E_{(x,y)}(k) . \eqn\Eneg $$  In the momentum space, the energy spectrum
$E_{(x,y)}(k)$
is given in terms of velocity correlators:
$$E_{(x,y)}(k) = {1\over 8\pi^{2}}
\int d^{2}x' e^{ik_{\alpha}(x_{\alpha}-x'_{\alpha})}\langle
v_{\beta}(x',y') v_{\beta}(x,y) \rangle_{b} .\eqn\ednk$$
For a large $k, \  E_{(x,y)}(k)$ can be obtained directly from Eqs.\onept \ \&
\opeII \
such that
$$\eqalign{ E_{(x,y)}(k) & \sim  \sum_{i} C_{\psi \psi }^{\phi_{i}}
k^{4 \Delta_{\psi } -2\Delta_{\phi_{i}}+1} \langle \phi_{i} \rangle_{b}
  +{\rm descendents}  \cr
& =   \sum_{i} C_{\psi \psi }^{\phi_{i}}k^{4 \Delta_{\psi }
-2\Delta_{\phi_{i}}+1} d_{\phi_{i}}y^{-2\Delta_{\phi _{i}}}
+{\rm descendents} .  }
\eqn\eneg$$
Thus we have obtained an explicit expression of the energy spectrum as a
function of the distance from the boundary. Unlike the bulk case, the
spectrum is made of a sum of different powers controlled by
$\Delta_{\phi _{i}}$ and $  d_{\phi_{i}} $.
The appearance of too many parameters, in general, is not desirable
in the statistical description of turbulence. However, if we move far away
from the boundary, i.e. when $y$ becomes very large, only the first few
terms in the series of Eq.\eneg \ dominate and the theory is controlled by
only few parameters. As we will discuss below, this is precisely
in agreement with the traditional Kolmogorov's idea of the inertial range.

 The main idea of Kolmogorov is first to assume that with sufficiently large
 Reynolds number, turbulence becomes locally isotropic -- i.e. small scale
 velocity field fluctuations are statistically homogeneous, isotropic, and
 stationary -- regardless of the form of a finite space-time region bounding
 turbulence.  Then Kolmogorov suggested  that for locally isotropic turbulence,
 the energy spectrum in the  intermediate scale (inertial range)  depends
 only on the constant flux of energy (or enstrophy for the two dimensional
 turbulence\refmark{\Kra }).
  In our case, we note that for sufficiently large $y$,
 conformal turbulence becomes locally isotropic. This is so because even in
 the presence of the anisotropic half plane boundary, the small scale
 fluctuations of velocity field become isotropic when $y$ becomes large.
 In order to see this explicitly, one could compare for example
 a small scale velocity fluctuations
  $\langle v_{x}(x, y)v_{x}(x,y+\epsilon )  \rangle_{b} $ with its $90^{\circ}$
 rotation   $\langle v_{y}(x, y)v_{y}(x + \epsilon,y )  \rangle_{b}; \epsilon
 \ll 1 $.
 In the leading order, they both agree to
 $$\langle v_{x}(x, y)v_{x}(x,y+\epsilon )  \rangle_{b} \approx
 \langle v_{y}(x, y)v_{y}(x + \epsilon,y )  \rangle_{b}  \approx
 -{2C_{\psi \psi }^{\phi }(2\Delta_{\psi }-\Delta_{\phi })d_{\phi }
 \over y^{2\Delta_{\phi }}\epsilon^{4\Delta_{\psi }-2\Delta_{\phi }+2}},
  \eqn \isot $$
 where   $\phi$ is the minimal dimension operator of  conformal dimension
$\Delta_{\phi}$ coming from the operator product $[\psi ][\psi ]$.
On the other hand, the constant enstrophy condition which follows
 from the $r$-independence of $\langle \dot{w}(r)w(0) \rangle$,
 can be met by a single condition when $y$ is large
  $$\Delta_{\phi} + \Delta_{\psi} - \Delta_{\chi} = -3 . \eqn\Enstro $$
where $\chi $ is the minimal dimenson operator coming from the OPE  of
$[\psi ][\phi ]$.\foot{
 The minimal dimensional operator $\chi $ from the OPE of  $[\psi ][\phi ]$
is not necessarily the same as the one coming from $[\psi ][\psi ][\psi ]$,
which were falsely identified in Ref.[3]. In many cases, they are equal but as
in the Table 1,  $M_{5,73} \ M_{5,77}$ are exceptions. In the case of
$M_{5,73}, \ \psi = \Phi_{(2,26)}, \  \phi = \Phi_{(1,15)}, \
\chi = \Phi_{(4,57)}$  and $\chi = [\phi ][\psi ]$ but
$\chi \ne [\psi ][\psi ][\psi ]$.}
In  case  $d_{\chi } = 0$, the constant
enstrophy condition changes into
  $$\Delta_{\phi} + \Delta_{\psi}  = -3 .  \eqn\Enst \ $$
Therefore, we contend that the inertial range forms in a distant region from
the boundary in the context of conformal turbulence while conformal
turbulence  generalizes turbulence to non-inertial regions. In general,
the inertial range develops regardless the shape of a particulrar boundary.
For example one point functions on a disk of radius $R$ can be obtained by
finite conformal transformation of Eq.\onept \ such that \refmark{\KyungheeII}
$$\langle\phi_i (r)\rangle_{\rm circle} =- {{(2R)^{2\Delta_{\phi_i}} d_{\phi_i}
} \over { (R^2-r^2)^{2 \Delta_{\phi_i}} }}.\eqn\opcir \ $$
This again shows that the minimal dimension operator dominates when $R$ becomes
large and $R\ll r$.

Having exemplified the role of one point functions,  we note that
there is no a priori reason to set parameters $d_{\phi_{i}}$ to zero.
It occurs only in the specific case of $SL(2,C)$ invariant vacuum or by
an unnatural fine tunning.
Thus, it is more natural to assume that $d_{\chi } \ne 0$ in our case
and use Eq.\Enstro \ for the constant enstrophy condition. This together
with the Navier-Stokes condition $\Delta_{\phi } > 2\Delta_{\psi }$
leads us to the new set of solutions of conformal turbulence on a upper
half plane which are listed in the Table 1.
In the numerical computation, we have used the following
notations and rules of 2-d CFT; $M_{p,q}$ denotes a minimal model
 with $p,q$ relatively prime and $p < q$, which is characterized by
${1\over 2}(p-1)(q-1)$ degenerate primary operators
$\Phi_{m,n} (1 \le n \leq q, \ \ 1 \le m \leq p )$ of conformal dimension
$$ \Delta_{m,n} = {(pn-qm)^{2} - (p-q)^{2} \over 4pq },\eqn\weight\ $$ and the
fusion
rules between two primaries;
$$\Phi_{m_{1},n_{1}} \times \Phi_{m_{2},n_{2}} = \sum_{i = |m_{1} - m_{2}|
+1 }^{r}\sum_{j=|n_{1}-n_{2}|+1}^{s}C^{(i,j)}_{(m_{1},n_{1})(m_{2},n_{2})}
\Phi_{i,j}, \eqn\fusionII\ $$ with $r =\min(m_{1}+m_{2}-1, 2p-m_{1}-m_{2}-1), \
s=\min(n_{1}+n_{2}-1, 2q-n_{1}-n_{2}-1)$, and $\
C^{(i,j)}_{(m_{1},n_{1})(m_{2},
n_{2})}$ are structure coefficients and $i,j$ run over odd or even numbers
if  it is bounded by  odd or even numbers respectively.

Finally, a couple of comments related to  new solutions are in order.

It is intriguing to observe that $\psi = \Phi_{(2,26)}$ occurs repeatedly
with increasing frequency as we change $p$ and $q$; $2 \le p \le 25 , \
p \le q \le 500 $ all of whose energy spectrum exponents lie in
between $ -4$ and $-5$.
Indeed, there exists a fixed point when $p \rightarrow \infty$ and
$q/p \rightarrow 15$.
In which case
$\psi = \Phi_{(2,26)}, \Delta_{\psi } = -3; \phi = \Phi_{(1,15)} ,
\Delta_{\phi } = - 49/15 ; \chi = \Phi_{(2,30)}, \Delta_{\chi } = -49/15 ,
$ and the energy spectrum exponent is $-67/15$. The physical relevance
of this fixed point is unknown.

 Table 1 shows that the energy spectrum exponents  can be greater than $-3$
 whereas in the bulk case they were strictly less than $-3$. This could be
 understood as a boundary effect to the energy spectrum. Different
 type of boundaries; e.g. different shapes (strip or disk) or periodic
 boundary conditions can also modify the one point functions and the
 energy spectrum. Details will appear in Ref.[12].

\noindent
{\bf Acknowledgements} \par
This work was supported in part by the program of
Basic Science Research, Ministry of Education,
and by Korea Science and Engineering Foundation,
and partly through CTP/SNU.  QP thanks K.Kang and Physics Department of Brown
University for their
support through the SNU-Brown exchange program during his visit.
\singlespace
\endpage
\refout
\endpage
{\centerline {Table 1: Some Solutions for
$\Delta_{\psi}+\Delta_{\phi}-\Delta_{\chi}=-3$}}
\vskip 1cm
\input tables
\begintable
$(p,q)$  | $\psi$ |  $\phi$  |  $\chi$  | ${\hbox {exponent}}$ ||
$(p,q)$  | $\psi$ |  $\phi$  |  $\chi$  | ${\hbox {exponent}}$ \crthick
$(2,33)$ | $(1,10)$   |  $(1,17)$ | $(1,16)$ | $-3.727273$         ||
$(3,43)$ | $(1,12)$   |  $(1,15)$ | $(1,14)$ | $-4.837209$          \cr
$(3,44)$ | $(2,26)$   |  $(1,15)$ | $(2,30)$ | $-4.613636$          ||
$(3,46)$ | $(2,26)$   |  $(1,15)$ | $(2,30)$ | $-4.282609$          \cr
$(3,52)$ | $(2,27)$   |  $(1,17)$ | $(2,35)$ | $-3.307692$          ||
$(3,67)$ | $(2,31)$   |  $(1,23)$ | $(2,45)$ | $-0.835821$          \cr
$(3,121)$| $(1,12)$   |  $(1,23)$ | $(1,34)$ | $-2.000000$          ||
$(3,169)$| $(1,14)$   |  $(1,27)$ | $(1,40)$ | $-2.000000$          \cr
$(3,196)$| $(1,15)$   |  $(1,29)$ | $(1,43)$ | $-2.000000$          ||
$(3,256)$| $(1,17)$   |  $(1,33)$ | $(1,49)$ | $-2.000000$          \cr
$(3,289)$| $(1,18)$   |  $(1,35)$ | $(1,52)$ | $-2.000000$          ||
$(3,361)$| $(1,20)$   |  $(1,39)$ | $(1,58)$ | $-2.000000$          \cr
$(3,400)$| $(1,21)$   |  $(1,41)$ | $(1,61)$ | $-2.000000$         ||
$(3,484)$| $(1,23)$   |  $(1,45)$ | $(1,67)$ | $-2.000000$          \cr
$(4,59)$ | $(2,26)$   |  $(1,15)$ | $(2,30)$ | $-4.580508$          ||
$(4,61)$ | $(2,26)$   |  $(1,15)$ | $(2,30)$ | $-4.331967$          \cr
$(4,75)$ | $(2,28)$   |  $(1,19)$ | $(2,38)$ | $-2.590000$          ||
$(4,95)$ | $(3,56)$   |  $(1,23)$ | $(3,72)$ | $-0.115789$          \cr
$(5,71)$ | $(4,55)$   |  $(1,15)$ | $(4,57)$ | $-4.929577$          ||
$(5,72)$ | $(4,55)$   |  $(1,15)$ | $(4,57)$ | $-4.777778$          \cr
$(5,73)$ | $(2,26)$   |  $(1,15)$ | $(2,30)$ | $-4.638356$          ||
$(5,74)$ | $(2,26)$   |  $(1,15)$ | $(2,30)$ | $-4.559459$          \cr
$(5,76)$ | $(2,26)$   |  $(1,15)$ | $(2,30)$ | $-4.360526$          ||
$(5,77)$ | $(2,26)$   |  $(1,15)$ | $(2,30)$ | $-4.241558$          \cr
$(5,91)$ | $(4,64)$   |  $(1,19)$ | $(4,72)$ | $-2.890110$          ||
$(5,106)$| $(2,30)$   |  $(1,21)$ | $(2,42)$ | $-1.371698$          \cr
$(5,114)$| $(4,77)$   |  $(1,23)$ | $(4,91)$ | $-0.578947$          ||
$(5,124)$| $(3,58)$   |  $(1,25)$ | $(3,74)$ | $ 0.424194$
\endtable

\end